\documentclass[fleqn]{annalen}
\usepackage{graphics}
\begin{document}
\newcommand{\volume}{8}              
\newcommand{\xyear}{1999}            
\newcommand{\issue}{5}               
\newcommand{\recdate}{29 July 1999}  
\newcommand{\revdate}{dd.mm.yyyy}    
\newcommand{\revnum}{0}              
\newcommand{\accdate}{dd.mm.yyyy}    
\newcommand{\coeditor}{ue}           
\newcommand{\firstpage}{1}         
\newcommand{\lastpage}{???}          
\def\xc{x_{\rm c}}
\def\pc{p_{\rm c}}
\def\pq{p_{\rm q}}
\setcounter{page}{\firstpage}        
\newcommand{\keywords}{quantum percolation, Anderson transition, universality} 
\newcommand{\PACS}{71.30.+h, 71.23.-k, 72.15.Rn}
\newcommand{\shorttitle}{A. Kaneko {\it et al.},
3D quantum percolation studied by level statistics}
\title{Three-dimensional quantum percolation studied by level statistics}
\author{Atsushi Kaneko and Tomi Ohtsuki} 
\newcommand{\address}
  {Department of Physics, Sophia University,
Kioicho 7-1, Tokyo 102-8554, JAPAN} 
\newcommand{\email}{\tt ohtsuki@sophia.ac.jp} 
\maketitle
\begin{abstract}
We study the metal-insulator transition on a three dimensional quantum 
percolation model by analyzing energy level statistics. 
The quantum percolation threshold $\pq$, which is 
larger than the classical percolation threshold $\pc$, 
becomes smaller when the time reversal symmetry (TRS) is broken,
i.e.  $\pq({\rm with \; TRS})>\pq({\rm without \; TRS})>\pc$.
It is shown that critical exponents are consistent
with the result of the Anderson transition, suggesting that
the quantum percolation problem
can be classified into the same universality classes of 
the Anderson transition. The shape of level statistics at the critical point 
is also reported.
\end{abstract}
\section{Introduction}
Due to its simple but non-trivial nature, percolation 
problems have been attracting a lot of attention \cite{stauffer}.
Especially interesting is its application to the 
transport properties in three-dinemsional disordered solids.
However, at low temperature, the quantum interference effects should
be seriously taken into account and the 
classical percolation picture is insufficient.
Thus the study of the quantum percolation problem becomes
a very important subject.

It is well known that disordered systems show the Anderson metal-
insulator transition.
This is classified into three universality classes 
(orthogonal, unitary, and symplectic universality classes) 
by time-reversal and spin-rotation symmetry. 
This conjecture is well confirmed for the Anderson tight binding
model \cite{SO}.
Then it is natural to ask whether the
quantum percolation problem can be also classified into the same 
universality classes of the Anderson transition.
If yes, the critical exponent $\nu$ for the divergence of the localization
length in the quantum percolation 
model is the same as in the Anderson model, namely,
$\nu=1.57\pm 0.02$ in the presence of the time reversal symmetry (TRS)
\cite{SO,SO2}, and $\nu=1.43\pm 0.02$ in the absence of it \cite{SO}.
However, the estimates of $\nu$ in the quantum percolation problem
reported so far are at variance.
For example, renormalization group analyses of the quantum percolation 
problem give $\nu=2.1$ \cite{OC,CO}, $\nu=1.9\pm 0.5$ \cite{RS}, 
and $\nu=1.86\pm0.02$ \cite{RBS}, well above the value in the Anderson
model.
From the Thouless number analysis, it is estimated to be
$\nu=1.95\pm0.12 $\cite{KN1,KN2}, again larger than in the
Anderson model.
On the other hand, the transfer matrix method for the network model gives
$\nu=0.75\pm0.1$\cite{AL}, and the analytic estimate of
the transmission coefficient gives
$\nu=0.38\pm 0.07$\cite{chang},
significantly smaller than in the Anderson model.
Recently, using the finite size scaling behavior of level statistics
\cite{SSSLS}, Berkovits and Avishai
estimated  $\nu$ to be $1.35\pm0.1$\cite{BA},
not far from the value of the Anderson model.
Inspired by their work, we perform extensive numerical study of the
level statistics on the percolation cluster.
We also study the effect of breaking of TRS.

\section{Model and method}
To describe the three dimensional (3D)
quantum bond percolation model, we consider the following
tight-binding Hamiltonian
\begin{equation}
H=\sum_{\langle ij \rangle}(t_{ij}a_i^{\dagger}a_j+{\rm H.c})
\end{equation}
where $\langle ij \rangle$ denotes nearest neighbors. The transfer 
energy is defined as 
\begin{equation}
t_{ij}=\cases{
        \exp({\rm i}\theta_{ij})   & (for connected bond) \cr
        0 & (for disconnected bond) \cr
}
\end{equation}
Bonds are randomly connected with probabilities $p$.
$\theta_{ij}$ is the Peierls phase due to magnetic field. 
When all the Peierls phases are set to 0, the Hamiltonian is time 
reversal symmetric, and we call it TRS model. 
When the phases are not vanishing, the Hamiltonian is generally 
not time reversal symmetric. We set $-\pi<\theta_{ij}<\pi$ randomly, 
and call this situation non-TRS model hereafter.
The underlying lattice is a three-dimensional cube of length 
$L$ with periodic boundary conditions. 
For each realization of bond structure, 
we first find maximally percolating cluster and then we  
 diagonalize it by Lanczos method. In order to gain good statistics,
more than $10^6$ eigenvalues are calculated,
so the number of realizations of random bond configuration are
$N=580,300,175$ and 110 for sample sizes 
$L^3=12^3,15^3,18^3$ and $21^3$. 
We take levels in 
the region $0.2<|E|<0.8$ where density of states 
is rather smooth \cite{BA}. 
Then the distribution function $P(s)$ of adjacent level spacings $s$
is calculated.
In the limit of large system size $L\to\infty$,
the level spacing distribution $P(s)$ 
is expected to be described by Poisson distribution 
$P(s)=\exp{(-s)}$ in the insulator region.
In the metallic regime, $P(s)$ is well described by 
the Wigner surmise, $P(s)\propto s^{\beta}\exp{(-A_{\beta}s^2)}$ 
($\beta=1$ in the presence of TRS and $\beta=2$, otherwise).

In order to obtain the critical value of the probability $\pq$ 
and critical exponent $\nu$, 
we define $I(s)$ and $\Lambda(p,L)$ as
\begin{equation}
\Lambda(p,L)= \frac{\int_0^{s_0} I(s){\rm d}s-\int_0^{s_0}
I_{\rm P}(s){\rm d}s}
{\int_0^{s_0} I_{\rm Wig}(s){\rm d}s-\int_0^{s_0} I_{\rm P}(s){\rm d}s},
\hspace{0.5cm}
I(s)=\int_0^sP(s'){\rm d}s'
\end{equation}
which characterizes the transition from Poisson to Wigner. 
$s_0$ is set to 1.2. 
Noting that $\xi(p)$  
diverges as
\begin{equation}
\xi(p)\sim |p-\pq|^{-\nu},
\end{equation}
near the critical probability $\pq$,  $\Lambda(p,L)$ is 
expected to behave as 
\begin{eqnarray}
\Lambda(p,L)=f[L/\xi(p)]=a_0+a_1[L(p-\pq)^{\nu}]^{1/\nu}+
a_2[L(p-\pq)^{\nu}]^{2/\nu}+\cdots.
\end{eqnarray}
Fitting the data to this expression, we estimate $\pq$ and $\nu$.

\section{Result}
The results are summarized in Table I.
When magnetic fields are applied, the critical point $\pq$ of non-TRS model
becomes smaller than that in the presence TRS, suggesting that
this transition has a nature of the Anderson transition. 
Note that $\pq$ is much larger than the classical
bond percolation threshold $\pc\approx 0.249$.
This is why taking all energy levels including small clusters \cite{BA}
and taking only those in the lagest cluster make small
difference \cite{avishai,kaneko}.
Both TRS and non-TRS models show 
$\nu$ consistent with the value of the Anderson
transition \cite{kaneko}.

We then compare the shape of level statistics 
at the critical point $\pq$(Fig. 1).
It is clearly seen that the critical level statistics is sensitive
to the breaking of TRS.
\begin{figure}
\centerline{\resizebox{11cm}{6cm}{\includegraphics{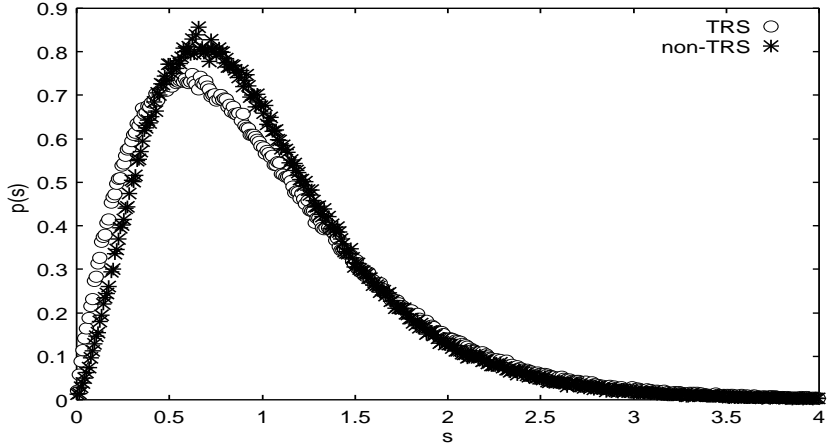}}}
\caption{Level spacing distribution for TRS (a) and non-TRS (b) models
at the critical point.
The system size is $L=21$ for example, but the form is
almost independet of $L$.}
\label{lsd}
\end{figure}

In the large $s$ region, the
level statistics is described by the sub-Poisson form,
$\propto\exp{(-\kappa s)}$ (Fig. 2).
In the Anderson model,
$\kappa =1.9\pm 0.1$ for the orthogonal symmetry \cite{ZK},
and $1.87$ for unitary symmetry \cite{BSZK}.
In the present quantum percolation model, $\kappa=1.77\pm 0.10$
in the presence of TRS, and for non-TRS model  
$\kappa=1.87\pm 0.03$.

\begin{figure}
\centerline{\resizebox{11cm}{6cm}{\includegraphics{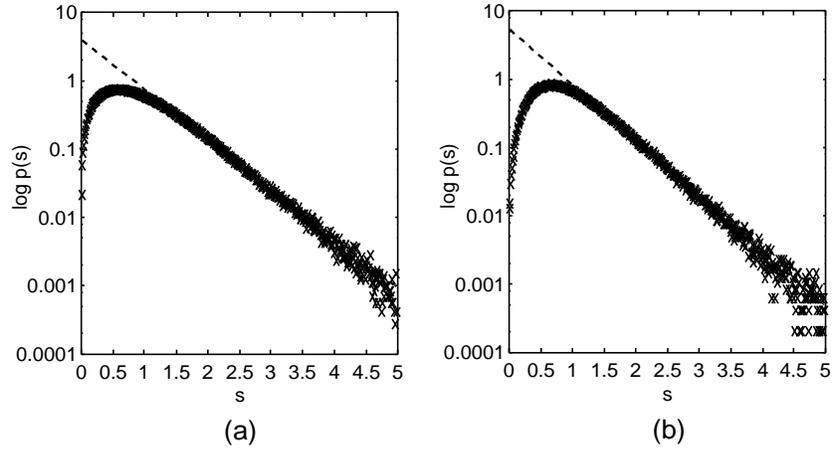}}}
\caption{Semi-log scale level spacing distribution for TRS(a) 
and non-TRS model(b).
We fit the curve to $C \exp{(-\kappa s)}$ with $\kappa=1.77$ for TRS model 
and $\kappa=1.87$ for non-TRS model.}
\label{log}
\end{figure}

The overall form of the critical level spacing distribution
deviates from those in the Anderson model with periodic
boundary condition (b.c.) and the fixed b.c. \cite{kaneko}.
Detailed will be reported elsewhere.

\begin{table}[h]
\caption{The critical point $\pq$ and the critical 
exponent $\nu$ with their standard deviations.
The values of the TRS model and the non-TRS model are only for
percolating cluster,  while TRS (all)
is obtained from all energy levels. 
TRS (all,BA) is the result reported in ref. \cite{BA}.}
\label{cp}
\vspace{0.2cm}
\begin{tabular}{@{\hspace{\tabcolsep}\extracolsep{\fill}}llll} \hline
system    & size                &    $\pq$           & $\nu$  \\ \hline
TRS (all,BA) & $L=7,9,11,13,15$ & $0.33\pm .01$      & $1.35\pm .10$ \\
TRS (all) & $L=12,15,18,21$     & $0.321\pm .001$    & $1.45\pm .07$ \\
TRS       & $L=12,15,18,21$     & $0.324\pm .001$    & $1.42\pm .07$ \\
TRS       & $L=15,18,21$        & $0.324\pm .001$    & $1.45\pm .11$ \\
\hline
non-TRS   & $L=12,15,18,21$     & $0.309\pm .001$    & $1.13\pm .05$ \\
non-TRS   & $L=15,18,21$        & $0.308\pm .001$    & $1.25\pm .08$ \\ 
\hline
\end{tabular}
\end{table}

%
%

\end{document}